\documentclass[aps,preprint,tightenlines,floatfix,superscriptaddress,nofootinbib]{revtex4}
\pdfoutput=1
\usepackage{graphicx}
\usepackage{dcolumn}
\def\beq{\begin{equation}}
\def\eeq{\end{equation}}
\def\bea{\begin{eqnarray}}
\def\eea{\end{eqnarray}}

\def\roughly#1{\mathrel{\raise.3ex\hbox
{$#1$\kern-.75em\lower1ex\hbox{$\sim$}}}}

\begin{document}
\bibliographystyle{apsrev}

\preprint{\vbox {\hbox{UdeM-GPP-TH-09-179}}}

\vspace*{2cm}

\title{\boldmath CP Violation in Three-Body Chargino Decays}

\def\umontreal{\affiliation{\it Physique des Particules,
Universit\'e de Montr\'eal, \\ C.P. 6128, succ. centre-ville,
Montr\'eal, QC, Canada H3C 3J7}}
\def\tayloru{\affiliation{\it Physics and Engineering Department, Taylor
University, \\ 236 West Reade Ave., Upland, IN 46989, USA}}

\umontreal
\tayloru

\author{Makiko Nagashima}
\email{makiko@LPS.umontreal.ca}
\umontreal

\author{Ken Kiers}
\email{knkiers@taylor.edu}
\tayloru

\author{Alejandro Szynkman}
\email{szynkman@lps.umontreal.ca}
\umontreal

\author{David London}
\email{london@lps.umontreal.ca}
\umontreal

\author{Jenna Hanchey}
\email{jenna.hanchey@gmail.com}
\tayloru

\author{Kevin Little}
\email{little@uchicago.edu}
\tayloru

\date{\today}

\begin{abstract}
CP violation in supersymmetry can give rise to rate asymmetries in the
decays of supersymmetric particles.  In this work we compute the rate
asymmetries for the three-body chargino decays ${\tilde\chi}^\pm_2 \to
{\tilde\chi}^\pm_1 HH$, ${\tilde\chi}^\pm_2 \to {\tilde\chi}^\pm_1
ZZ$, ${\tilde\chi}^\pm_2 \to {\tilde\chi}^\pm_1 W^+ W^-$ and
${\tilde\chi}^\pm_2 \to {\tilde\chi}^\pm_1 ZH$.  Each of the decays
contains contributions mediated by neutral Higgs bosons that can
possibly go on shell.  Such contributions receive a resonant
enhancement; furthermore, the strong phases required for the CP
asymmetries come from the widths of the exchanged Higgs bosons.  Our
results indicate that the rate asymmetries can be relatively large in
some cases, while still respecting a number of important low-energy
bounds such as those coming from $B$ meson observables and electric
dipole moments.  For the parameters that we consider, rate asymmetries
of order $10\%$ are possible in some cases.
\end{abstract}


\maketitle

\newpage
\setcounter{page}{1}

Supersymmetry (SUSY) has been proposed as a solution to the hierarchy
problem in the standard model (SM). SUSY is widely thought to be the
physics that lies beyond the SM, and it is hoped that it will be
discovered in the future at the LHC or at a linear collider.

In SUSY theories each ordinary fermion and gauge boson has a
superpartner, respectively of spin 0 and spin $\frac12$.  SUSY also
includes a charged Higgs boson, the $H^-$. The $W^-$ and $H^-$ each
have a fermionic partner, known as charginos. These two charginos can
mix, resulting in two mass eigenstates ${\tilde\chi}^-_1$ and ${\tilde
\chi}^-_2$ whose masses can be very different. Here we adopt the
standard notation $m_{{\tilde \chi}^-_2} > m_{{\tilde \chi}^-_1}$. In
this paper, we study CP violation in the decay ${\tilde\chi}^\pm_2 \to
{\tilde\chi}^\pm_1 XY$, where $XY = HH$, $ZZ$, $W^+W^-$ or $ZH$ ($H$
is a Higgs boson).  This analysis is an extension of the work
performed in Refs~\cite{previous1,previous2}.  Throughout this paper
we assume that the CP-conserving SUSY parameters are known, but that
those which violate CP remain to be measured.  Some previous studies
of CP violation involving charginos may be found in
Refs.~\cite{yang,bartl2004,kittel,eberl,bartl2006,osland,rolbiecki,bartl,Kittel:2008be}.

CP violation in SUSY has been studied extensively at low
energies, in meson mixing \cite{mixing}, in the $B$-meson
system \cite{B-meson} and in electric dipole moments (EDMs)
\cite{edms}. EDMs in particular provide quite stringent
constraints on the low-energy CP-violating SUSY phases of the
superparticle couplings.  For certain values of the SUSY
parameters, there is a disagreement with the experimental
limits, resulting in the so-called SUSY CP problem.  In this
paper we make extensive use of the computer program {\tt
CPsuperH2.0} \cite{CPsuperH,CPsuperH2a,CPsuperH2b} in choosing
values for the various SUSY parameters in our processes.  The
most recent version of this program allows the user to
compute EDMs, as well as various other low-energy
observables.

In a given decay, there are three types of CP-violating signals.
First, there is the partial rate asymmetry.  Any difference in the
rate between process and CP-conjugate process is a signal of CP
violation.  As we will see, it is possible that the partial rate
asymmetry is sizeable (of order 5-10\% for the processes considered).
Second, one has the modified (spin-dependent) rate asymmetry.  Here
one compares the rate for process and CP-conjugate process for the
case in which the spin of one of the particles has been measured.
This measurement is complicated, and so, given that the partial rate
asymmetry can be significant, we do not consider the modified rate
asymmetry in this paper.  The third CP-violating signal is the
triple-product (TP) asymmetry, which is proportional to $\vec v_1
\cdot (\vec v_2 \times \vec v_3)$ (each $v_i$ is a spin or momentum).
This is due to terms of the form ${\rm Tr} [\gamma_\alpha \gamma_\beta
  \gamma_\rho \gamma_\sigma \gamma_5]$ in the square of the amplitude.
Since only 3-body decays are considered here, a nonzero TP can arise
only if a spin or polarization is measured.  As noted above, such
measurements are difficult, and so we do not consider TPs here.  Thus,
the only CP-violating signal analyzed in this paper is the partial rate
asymmetry.

All effects that violate CP require the interference of (at
least) two amplitudes.  The partial rate asymmetry is
proportional to $\sin\delta$, where $\delta$ is the relative
strong (CP-even) phase between the interfering amplitudes.
Strong phases can be generated in one of two ways. First, one
can have the exchange of gluons between the particles
involved in the decay, leading to QCD-based strong
phases. Unfortunately, we do not know how to calculate the
strong phases in this case. Alternatively, the strong phases
can be generated by the (known) widths of the intermediate
particles in the decay.  In the decays considered in this
paper, the particles do not couple to gluons. Thus, the
strong phase arises only due to the widths of the
intermediate particles.  This is good, given that we want to
{\it measure} the CP violation, and not simply detect its
presence.

Now, SUSY includes two Higgs doublets which contain (in the
gauge basis) two neutral scalars and one pseudoscalar. In the
mass basis, these particles mix and one obtains three mass
eigenstates $H_1$, $H_2$ and $H_3$. In SUSY theories, the
lightest mass is $m_{H_1} = O(100)$ GeV and we therefore take
the final-state $H$ to be $H_1$.

\begin{figure}[t]
\begin{center}
\resizebox{5in}{!}{\includegraphics*{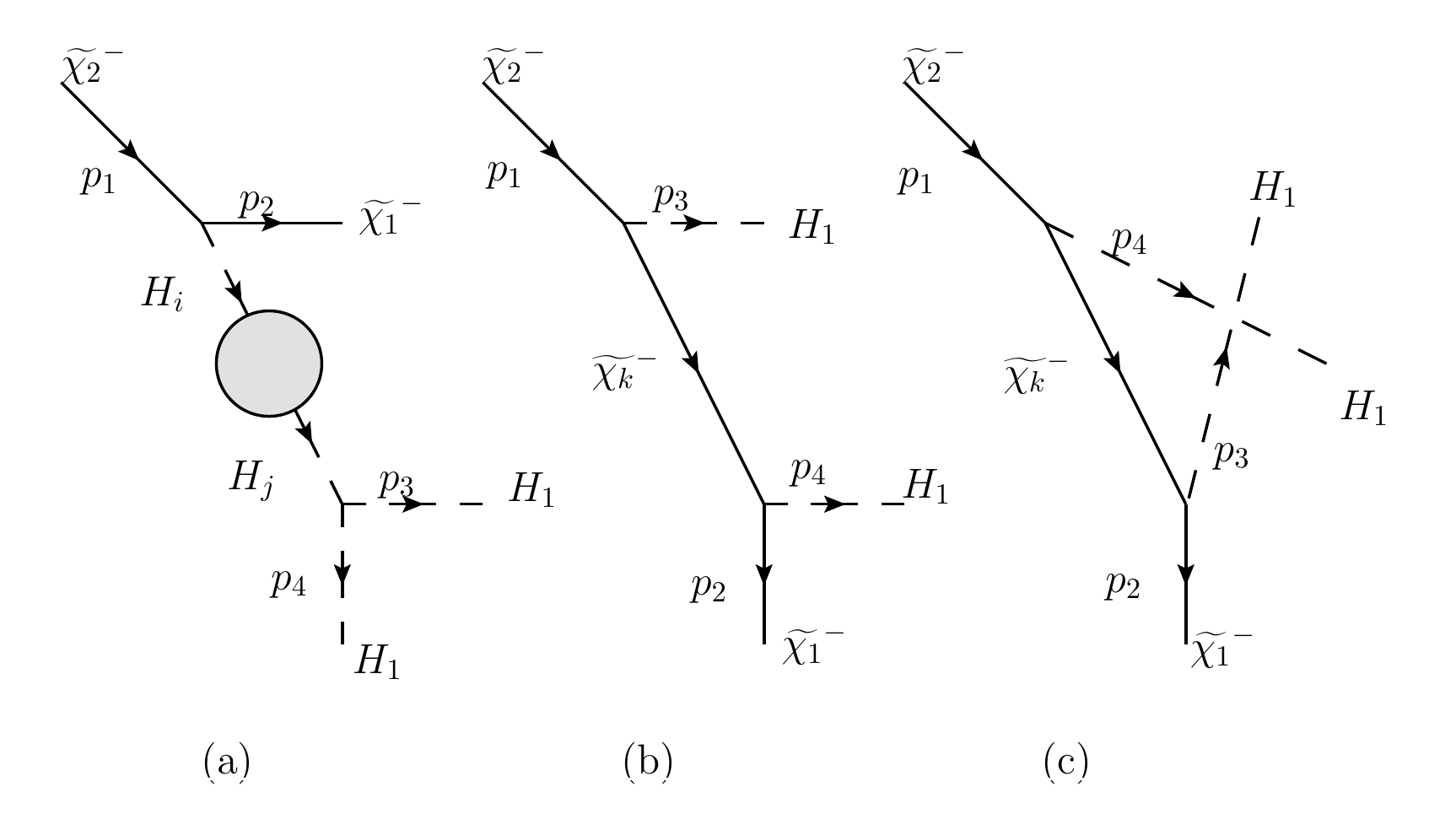}}
\caption{Feynman diagrams for the decay ${\tilde\chi}^-_2 \to
{\tilde\chi}^-_1 H_1H_1$. In diagram (a) the decay is mediated by
neutral Higgs bosons, while diagrams (b) and (c) are mediated by
charginos.}
\label{fig:feyn_diag1}
\end{center}
\end{figure}

Figure~\ref{fig:feyn_diag1} shows the diagrams that contribute to
${\tilde\chi}^-_2 \to {\tilde\chi}^-_1 H_1H_1$.  The diagrams for the
$ZZ$ case are identical, but with $H_1$ replaced by $Z$ in the final
state.  The $WW$ diagrams are similar, except that there is no diagram
analogous to Fig.~\ref{fig:feyn_diag1} (b) (we take ``$p_4$'' to
correspond to the $W^-$ for this decay).  Furthermore, the diagram
analogous to Fig.~\ref{fig:feyn_diag1} (c) involves intermediate
neutralinos instead of charginos, and there is an extra diagram
similar to Fig.~\ref{fig:feyn_diag1} (a), but with the intermediate
Higgs bosons replaced by the $Z$.  Finally, for the $ZH_1$ case the
diagrams are similar to Fig.~\ref{fig:feyn_diag1}, but with one $H_1$
replaced by $Z$.  Also, there is an extra diagram similar to
Fig.~\ref{fig:feyn_diag1} (a), but with the intermediate Higgs bosons
replaced by the $Z$.  Note that the neutralinos, ${\tilde\chi}^0_i$
($i=1,\ldots 4$), are the fermionic partners of the $\gamma$, $Z$, and
neutral Higgs bosons.

The decay amplitudes that are of most interest to us arise from
diagrams containing an intermediate particle that can go
on-shell. Such diagrams can benefit from a resonant enhancement.  The
internal ${\tilde\chi}^-_{1,2}$, $H_1$ and $Z$ can never be on-shell,
while the $H_2$ and $H_3$ can be. (For a given set of SUSY parameters,
the widths $\Gamma_2$ and $\Gamma_3$ are calculable, as are the
``off-diagonal widths'' associated with transitions
$H_i\leftrightarrow H_j$ \cite{widths}.  These terms are taken into
account by {\tt CPsuperH2.0} when it computes the neutral Higgs
propagator matrix.)  The case of internal neutralinos (which appear in
the case $XY=W^+W^-$) is more complicated. In principle, these could
be on-shell.  Unfortunately, we do not know their widths, making their
contributions to the rate asymmetry uncertain.  In practice, for the
cases that we consider, the two heavier neutralinos have masses close
to $m_{{\tilde \chi}^-_2}$ and the two lighter neutralinos have masses
close to or less than $m_{{\tilde \chi}^-_1}$.  Thus, in the examples
we consider, the neutralinos cannot go on shell and the uncertainty
associated with the neutralinos' (unknown) widths is mitigated.

In order for a given partial rate asymmetry to be appreciable, the
interfering amplitudes should be of comparable sizes.
Furthermore, since the rate asymmetries depend on an
integration over phase space, the ``large'' contributions
from these amplitudes should occur in similar regions of
phase space.  This latter requirement is met if the masses of
the on-shell particles, $H_2$ and $H_3$, are similar.
Fortunately, it is relatively common in SUSY theories that
$m_{H_2} \simeq m_{H_3}$~\cite{higgs_tree}.  The decay
amplitudes of most interest to us, therefore, are those
dominated by internal $H_2$ and $H_3$ exchange.  In
calculating the square of the amplitude, however, we retain
the contributions of various non-resonant diagrams (see
Fig.~\ref{fig:feyn_diag1}), since they can in principle give
non-negligible contributions.  With these ingredients, we
calculate the partial rate asymmetry. Since it is assumed that
the masses are known, we can determine whether or not CP
violation is likely in these decays.  Measurement of these CP
asymmetries will allow experimentalists to extract and/or
constrain the SUSY parameters, including CP-violating
SUSY phases.

\begin{figure}[t]
\begin{center}
\resizebox{5.8in}{!}{\includegraphics*{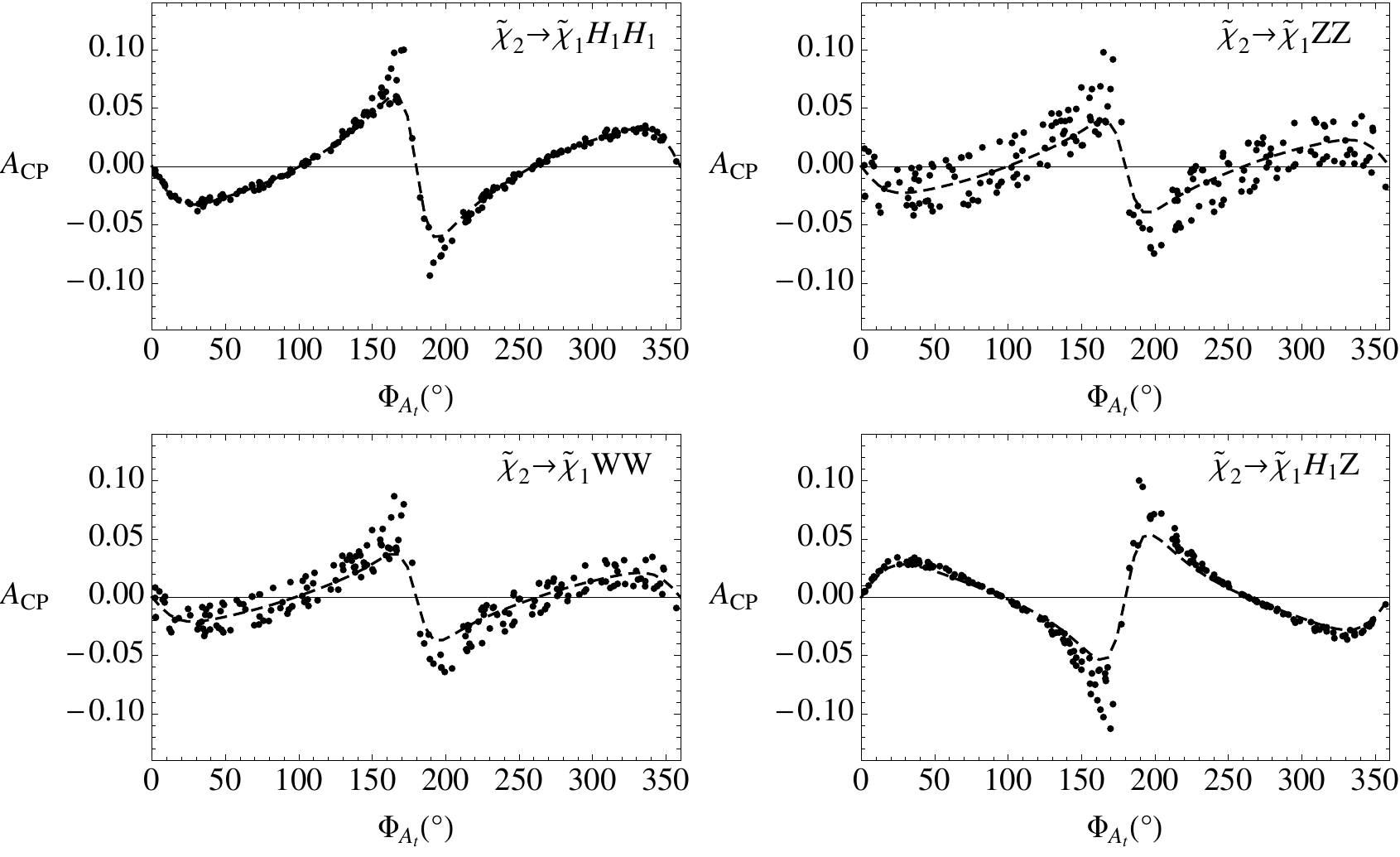}}
\caption{Plots of partial rate asymmetries as a function of $\Phi_{A_t}$.  The
  dashed lines show the asymmetries for the case
  $\Phi_1$$=$$\Phi_2$$=$$\Phi_3$$=$$\Phi_{A_b}$$=$$\Phi_{A_\tau}$$=0^\circ$
  (the values of the other SUSY parameters are given in the text).
  The scattered points indicate the asymmetries obtained by allowing
  all six phases to vary randomly between $0^\circ$ and $360^\circ$.
  Low-energy experimental results have not been used to constrain the
  SUSY parameter space when generating these plots.}
\label{fig:acp_fn_at}
\end{center}
\end{figure}

\begin{figure}[t]
\begin{center}
\resizebox{5.8in}{!}{\includegraphics*{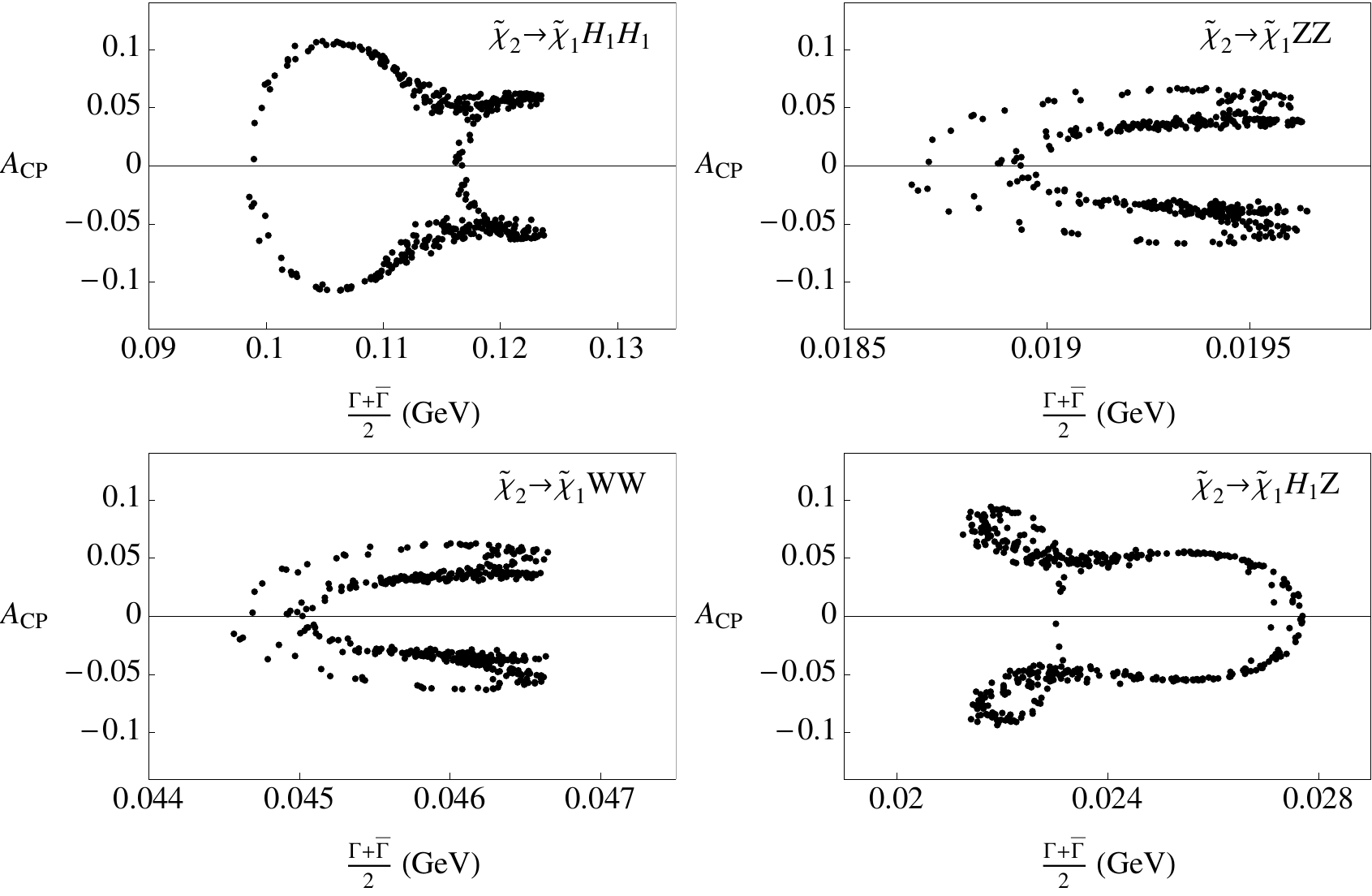}}
\caption{Scatter plots of partial rate asymmetries versus mean widths for
  ${\tilde\chi}_2 \to {\tilde\chi}_1 XY$, with $XY = H_1H_1$,
  $ZZ$, $W^+W^-$ or $ZH_1$.}
\label{fig:scatterplots}
\end{center}
\end{figure}

Analytical expressions for the amplitudes in question may be
found in the Appendix.  Our notation is similar or identical
to that employed in Ref.~\cite{CPsuperH}.  We have also used
Feynman rules derived from Refs.~\cite{rosiek1,rosiek2} where
necessary.  FeynCalc~\cite{feyncalc} was used to compute the
squares of the amplitudes.  The resulting expressions are
quite messy and have not been included here.  Please also
note the following:
\begin{enumerate}

\item Although only $H_2$ and $H_3$ can go on-shell,\footnote{As noted
  above, the neutralinos can in principle go on-shell in the ``$WW$''
  case, but they do not do so for the parameters that we use.} we have
  allowed for the possibility of non-zero widths for other
  intermediate particles.  For the purpose of our numerical work, all
  non-Higgs intermediate particles (with the exception of the $Z$,
  whose width is known) have been given a uniform width of 10~GeV.
  Since the intermediate particles in question are off-shell, the
  quantities that we compute should not be very sensitive to the
  value(s) assumed for the intermediate particles' widths.  We have
  explicitly checked the effect of changing the uniform width from 10
  GeV to 1 GeV and to 20 GeV for the data points plotted in
  Figs.~\ref{fig:acp_fn_at} and \ref{fig:scatterplots} and have
  confirmed that the quantities shown in these plots are not very
  sensitive to such changes.
\item The inclusion of intermediate particles' widths in our
  calculation implies that certain beyond-tree-level diagrams
  have effectively been taken into account.  Other such
  diagrams have not been included, leading to possible issues
  with gauge dependence or with respect to the CP-CPT 
  connection~\cite{gerardhou,wolfenstein}.  {\tt
  CPsuperH2.0} uses the Pinch Technique when computing the
  elements of the neutral Higgs propagator
  matrix~\cite{pinch1,pinch2}.  This technique was designed
  in such a way that certain classes of diagrams would be
  gauge independent.  The neutral Higgs propagator matrix is
  a $4\times4$ matrix in this approach rather than the
  $3\times3$ matrix that one might expect.  Rigorous
  application of the Pinch Technique for the present
  calculation is beyond the scope of our paper.  Instead, we
  have used the $3\times3$ physical Higgs block of the
  $4\times4$ propagator computed by {\tt CPsuperH2.0} and have
  used Feynman rules consistent with the Unitary gauge.
\item It is understood that $H_1$ is not, in general, a CP-eigenstate,
  since the scalar and pseudoscalar Higgs bosons can mix when CP is
  violated.  Nevertheless, we have checked numerically that the
  asymmetries for the $H_1H_1$ and $ZH_1$ case are in fact zero when
  all CP-violating phases are zero.

\end{enumerate}

Having used the expressions in the Appendix to compute the
widths for the processes and CP-conjugate processes, we
subtract one from the other to obtain the respective partial rate
asymmetries,
\begin{eqnarray}
  {\cal A}_{\mbox{\scriptsize CP}}^{XY} \equiv 
  \frac{\Gamma({\tilde\chi}^-_2 \to {\tilde\chi}^-_1 XY)
    -\Gamma({\tilde\chi}^+_2 \to {\tilde\chi}^+_1 XY)}
  {\Gamma({\tilde\chi}^-_2 \to {\tilde\chi}^-_1 XY)
    +\Gamma({\tilde\chi}^+_2 \to {\tilde\chi}^+_1 XY)}\; ,
\end{eqnarray}
with $XY = H_1H_1$, $ZZ$, $W^+W^-$ or $ZH_1$.

The SUSY parameter space to be considered is enormous, and we have not
made any attempt to perform a systematic parameter scan in our
numerical work.  Rather, we have chosen to focus on a small region of
parameter space that is of interest.  In particular, we have focused
on a variation of the ``CPX'' scenario described in
Ref.~\cite{CPsuperH2b}.  In the CPX scenario, the SUSY parameters
$\mu$, $A_t$, $A_b$ and $A_\tau$ are chosen in such a way that
$\left|\mu A_{t,b,\tau}\right|=8M_{\mbox{\scriptsize SUSY}}^2$, where
$M_{\mbox{\scriptsize SUSY}}$ represents the common mass scale of the
third-generation squarks and sleptons (see
Refs.~\cite{CPsuperH2b,CPXoriginal} for further details),
and $A_{t,b,\tau}$ are trilinear couplings in the soft SUSY-breaking
Lagrangian.  This
scenario was originally motivated by the observation that certain
CP-violating terms in the neutral Higgs mass-squared matrix depend on
$\mbox{Im}\left(\mu A_t\right)/M_{\mbox{\scriptsize SUSY}}^2$.  In our
variation on the CPX scenario, we set $\left|\mu\right|=0.6$~TeV and
$M_{{\tilde Q}_3}$$=$$M_{{\tilde U}_3}$$=$$M_{{\tilde
    D}_3}$$=$$M_{{\tilde L}_3}$$=$$M_{{\tilde
    E}_3}$$=$$M_{\mbox{\scriptsize SUSY}}$$=$$0.5$~TeV, as well as
$\left|M_2\right|=2\left|M_1\right|=200$~GeV,
$\left|M_3\right|=1$~TeV, $m_{H^\pm}=300$~GeV and $\tan\beta=5$.
Also, we set
$\left|A_{u,d,c,s,e,\mu}\right|$$=$$\left|A_{t,b,\tau}\right|=1$~TeV
and $\Phi_{A_{u,d,c,s,e,\mu}}=0^\circ$.  Our notation here is the same
as that used in Ref.~\cite{CPsuperH2b}.

Figure~\ref{fig:acp_fn_at} shows the four asymmetries under
consideration, plotted as functions of $\Phi_{A_t}$.  For the purpose
of the plots, the hierarchy factors $\rho_{\tilde{J}}$ ($J=Q,
U,D,L,E$) have all been set to 10.\footnote{In the notation of {\tt
    CPsuperH2.0}, the mass parameters $M_{\tilde{J}_{1}}$ and
  $M_{\tilde{J}_{2}}$ are assumed to be equal to each other (here
  $J=Q, U,D,L,E$ and the subscripts ``1'' and ``2'' refer to the first
  and second generation, respectively).  The third generation mass
  parameters are allowed to be different from those of the first two.
  The hierarchy factors are defined via the following expression:
  $M_{\tilde{J}_{1,2}}=\rho_{\tilde{J}}M_{\tilde{J}_{3}}$.}  The
dashed line in each plot shows the result obtained by allowing
$\Phi_{A_t}$ to vary between $0^\circ$ and $360^\circ$, while keeping
the other five phases set to zero. (I.e., we set
$\Phi_1$$=$$\Phi_2$$=$$\Phi_3$$=$$\Phi_{A_b}$$=$$\Phi_{A_\tau}$$=0^\circ$,
where $\Phi_{1,2,3}$ are the phases associated with the complex
gaugino mass parameters $M_{1,2,3}$.  Note that throughout this work
we adopt the convention that $\Phi_\mu=0^\circ$.)  The scattered points in
these plots show the asymmetries obtained by allowing all six phases
to vary randomly between $0^\circ$ and 360$^\circ$.  As is evident
from the figure, for the parameters we have chosen, the asymmetries
are strongly dependent on $\Phi_{A_t}$, although other phases
contribute to the asymmetries as well.  Analogous plots, showing the
asymmetries as functions of the other five phases, do not demonstrate
the same pronounced dependence on the other phases.

\begin{table}
\caption{Constraints imposed when choosing SUSY parameter values for
  Figs.~\ref{fig:scatterplots} and \ref{fig:phasecorr}.  The first
  five rows refer to the EDMs for Thallium, the electron, Mercury, the
  neutron and the muon, respectively; the sixth row contains the bound
  we enforce for the SUSY contribution to the muon anomalous magnetic
  dipole moment.  Further discussion of some of the constraints may be
  found in the text.  Note the following: (1) there is some variation
  in the confidence levels corresponding to the experimental upper
  bounds quoted in the second column; (2) Refs.~\cite{CPsuperH2b} and
  \cite{ginges2003} contain further information regarding the Thallium
  EDM; and (3) to obtain the bound for the $B\to X_s\gamma$ branching
  ratio we have combined errors in quadrature to obtain $(3.52\pm
  0.25)\times10^{-4}$ and have doubled the uncertainty.  The range
  quoted in the table, and the constraint imposed, is thus at
  approximately the $2\sigma$ level.}
\begin{tabular}{|ccc|}
	\hline\hline
~~~~~~~Quantity~~~~~~~  & ~~~~~Constraint Imposed~~~~~   
  & ~~~References~~~    \\
	\hline \hline
$\left|d_{\mbox{\scriptsize Tl}}\right|$
        & $<9\times10^{-25}~e$~cm 
        & \cite{CPsuperH2b,regan2002,ginges2003}\\
$\left|d_e\right|$
        & $<1.6\times10^{-27}~e$~cm 
        & \cite{regan2002}\\
$\left|d_{\mbox{\scriptsize $^{199}$Hg}}\right|$
        & $<3.1\times10^{-29}~e$~cm 
        & \cite{griffith2009}\\
$\left|d_n\right|$
        & $<2.9\times10^{-26}~e$~cm 
        & \cite{baker2006}\\
	\hline
$\left|d_\mu\right|$
        & $<1.8\times10^{-19}~e$~cm 
        & \cite{bennett}\\
$a_\mu^{\mbox{\scriptsize SUSY}}$ 
        & $(19.45\pm19.45)\times10^{-10}$
        & \cite{cheung2009}\\
	\hline
${\cal B}(B\to X_s\gamma)$ 
        & $\left(3.52\pm 0.50\right)\times 10^{-4}$
        & \cite{hfag}\\
${\cal A}_{CP}(B\to X_s\gamma)$ 
        & $-0.012\pm 0.028$ 
        & \cite{hfag}\\
${\cal B}(B_s\to\mu^+\mu^-)$ 
        & $<4.7\times10^{-8}$ 
        & \cite{pdg}\\
${\cal B}(B_d\to\tau^+\tau^-)$ 
        & $<4.1\times10^{-3}$ 
        & \cite{pdg}\\
\hline
$\left|\,\Delta M_{B_d}^{\mbox{\scriptsize SUSY}}\right|$ & $<0.005$~ps$^{-1}$ & \cite{pdg} \\
$\left|\Delta M_{B_s}^{\mbox{\scriptsize SUSY}}\right|$ & $<0.12$~ps$^{-1}$ & \cite{pdg} \\
\hline\hline
\end{tabular}
\label{tab:table1}
\end{table}

In generating the data for Fig.~\ref{fig:acp_fn_at} we have not made
any attempt to impose the various low-energy experimental constraints
that are available, since the purpose of the plots is to demonstrate
the functional dependence of the asymmetries on the phases.  We now
turn to a more careful consideration of the asymmetries by also taking
into account several low-energy constraints.

The constraints we impose are listed in Table~\ref{tab:table1}, which
also contains some comments regarding the constraints.  We offer here
a few additional comments.  Let us first consider the muon anomalous
magnetic moment.  According to the authors of Ref.~\cite{cheung2009},
the experimental value for the muon anomalous magnetic moment exceeds
the SM prediction by $\Delta a_\mu$$=$$a_\mu^{\mbox{\scriptsize
    exp}}-a_\mu^{\mbox{\scriptsize SM}}$$=$$\left(30.7\pm
8.2\right)\times 10^{-10}$, which represents a $3.7\sigma$ deviation.
One option would be to require that the SUSY contribution make up the
difference between the experimental value and the SM prediction.  We
take a somewhat broader view and require the SUSY contribution to be
between zero and $(30.7+8.2)\times 10^{-10}=38.9\times10^{-10}$.  The
constraint quoted in Table~\ref{tab:table1} is thus
$a_\mu^{\mbox{\scriptsize SUSY}}= (19.45\pm19.45)\times10^{-10}$.  The
last two rows of the table describe constraints on the SUSY
contributions to $\Delta M_{B_d}$ and $\Delta M_{B_s}$.  The
experimental values for these quantities are $\Delta
M_{B_d}=0.507\pm0.005$~ps$^{-1}$ and $\Delta M_{B_s}=17.77\pm
0.12$~ps$^{-1}$~\cite{pdg}.  The corresponding constraints that we
have listed are thus just the experimental uncertainties in these
quantities.  These constraints are tighter than they need to be, since
we are ignoring the theoretical uncertainties within the SM.
Nevertheless, these particular constraints are easily passed for the
parameters we consider.  One constraint that we have not directly
imposed is on the ratio $R_{B\tau\nu}\equiv {\cal
  B}(B^-\to\tau^-\overline{\nu})/ {\cal B}^{\mbox{\scriptsize
    SM}}(B^-\to\tau^-\overline{\nu})$.  There has been some discussion
in the literature regarding the possible range of this ratio.  For the
parameters used to generate Fig.~\ref{fig:scatterplots} (see below),
we find $R_{B\tau\nu}\sim 0.985$, which is easily within the range
derived, for example, in Ref.~\cite{ellis2007}.

Figure~\ref{fig:scatterplots} shows the four asymmetries plotted as
functions of the mean width (defined to be
$\left[\Gamma({\tilde\chi}^-_2 \to {\tilde\chi}^-_1 XY)
  +\Gamma({\tilde\chi}^+_2 \to {\tilde\chi}^+_1 XY)\right]/2$, with
$XY = H_1H_1$, $ZZ$, $W^+W^-$ or $ZH_1$).  The SUSY parameter values
or ranges used to generate the plots in Fig.~\ref{fig:scatterplots}
are the same as those used to generate the scattered points in
Fig.~\ref{fig:acp_fn_at}, but with three changes.  First, instead of
fixing the $\rho_{\tilde{J}}$ ($J=Q, U,D,L,E$) to a particular value,
as was done for Fig.~\ref{fig:acp_fn_at}, we have now allowed the five
hierarchy parameters to vary independently (and randomly) in the range
$8$ to $12$.  Second, we have imposed the constraints described in
Table~\ref{tab:table1}.  (The quantities in the table are computed
automatically by the {\tt CPsuperH2.0} software.)  Imposing the
constraints from Table~\ref{tab:table1} leads to relatively strong
limits on the SUSY parameter space.  This led us to make a third
change: to increase the efficiency of our numerical work, we allowed
$\Phi_2$ only to take on the values $0^\circ$ and $180^\circ$, and we
restricted $\Phi_{A_t}$ so that it took on values between $140^\circ$
and $220^\circ$.  Of the $10^4$ parameter sets that were originally
generated in this manner, 432 were able to pass all of the cuts.  The
432 parameter sets did not contain any cases in which $\Phi_2$ was
$180^\circ$ (i.e., only the parameter sets with $\Phi_2=0^\circ$
survived the cuts).  Parameter sets that passed all of the constraints
were used to compute the widths and asymmetries that appear in
Fig.~\ref{fig:scatterplots}.\footnote{A few technical notes regarding
  our calculation are the following: (i) {\tt CPsuperH2.0} follows
  three separate approaches when computing the neutron EDM and thus
  gives three separate estimates for the EDM~\cite{CPsuperH2b}.  In
  our numerical work, we insisted that at least one of these three
  numbers satisfy the neutron EDM constraint listed in
  Table~\ref{tab:table1}. (ii) We used the (approximate) default
  method for dealing with the dimension-six Weinberg operator, rather
  than supplying our own integration routine.  See
  Ref.~\cite{CPsuperH2b} for further discussion.}  For the sets of
parameters considered, and respecting various low-energy bounds,
asymmetries of order 10\% are possible for ${\tilde\chi}_2 \to
{\tilde\chi}_1 H_1H_1$ and slightly smaller asymmetries occur for the
other decay modes. 

With the parameters used for the plots in Fig.~\ref{fig:scatterplots},
the three neutral Higgs bosons had masses $m_{H_1}\sim
112.9-119.5$~GeV, $m_{H_2}\sim 289.3-290.3$~GeV and $m_{H_3}\sim
291.1-292.1$~GeV.  Also, the lighter and heavier charginos had masses
$m_{\widetilde{\chi_1}}\sim 191.6$~GeV and $m_{\widetilde{\chi_2}}\sim
613.3$~GeV, respectively.  Thus, the parameters were such that the two
heavier Higgs bosons could go on shell when mediating the various
decay processes, as was noted above.

\begin{figure}[t]
\begin{center}
    \resizebox{3in}{!}{\includegraphics*{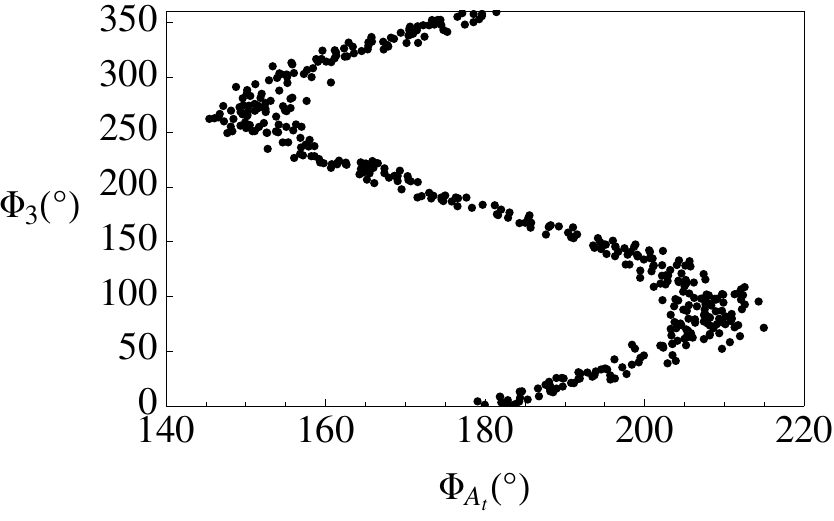}}
\caption{Correlations between $\Phi_3$ and $\Phi_{A_t}$.  The points
  shown correspond to the same parameters used to generate
  Fig.~\ref{fig:scatterplots}.}
\label{fig:phasecorr}
\end{center}
\end{figure}

Figure~\ref{fig:phasecorr} shows the correlations between $\Phi_3$ and
$\Phi_{A_t}$ for the parameter sets that passed the constraints from
Table~\ref{tab:table1} and that were subsequently used for the plots
in Fig.~\ref{fig:scatterplots}.  As is evident from the figure, there
is a strong correlation between these two phases that comes into play
in allowing the constraints to be passed.

It is useful to consider the observability of a 10\% partial rate
asymmetry.  After considering the main contributions from the open
two-body decay channels, we estimate the total width of the heavier
chargino to be of order $10$$-$$20$ GeV within the allowed parameter
space region. This result, together with the ${\tilde\chi}_2 \to
{\tilde\chi}_1 H_1H_1$ partial width, allows us to obtain a
representative value for the statistical significance of the CP
asymmetry, $S \sim |A_{CP}| \sqrt{2N}$, where $N$ is the number of
events corresponding to the decay under study.  ($N$ represents the
number of $\tilde\chi_2^+$ events, as well as the number of
$\tilde\chi_2^-$ events; these are assumed to be similar.)  If, for
example, the partial rate asymmetries are measured at an $e^+ e^-$
linear collider, with a single ${\tilde\chi}_2$ production cross
section of order 20 fb and an integrated luminosity of $\mathcal{L} =
500$ $\mbox{fb}^{-1}$ \cite{bartl2004,kittel}, then the statistical
significance turns out to be of order unity.  This value should be
understood as a very crude estimate, since the ${\tilde\chi}_2$
production cross section is strongly dependent on the SUSY parameters
and since we have not preformed a detailed analysis, nor considered
the case of the LHC.  Our point is simply to show that a 10\% partial
rate asymmetry could well be reachable.

In conclusion, if SUSY is discovered in future experiments, it will
become important to measure the underlying parameters of the theory.
This paper has examined four chargino decay modes that could help
provide insight into the CP nature of the theory.  In particular, we
have computed partial rate asymmetries for the decays ${\tilde\chi}^\pm_2 \to
{\tilde\chi}^\pm_1 H_1H_1$, ${\tilde\chi}^\pm_2 \to {\tilde\chi}^\pm_1
ZZ$, ${\tilde\chi}^\pm_2 \to {\tilde\chi}^\pm_1 W^+ W^-$ and
${\tilde\chi}^\pm_2 \to {\tilde\chi}^\pm_1 ZH_1$.  Rate asymmetries
have an advantage over some other CP-violating observables in that no
spins or polarizations need to be measured.  In the numerical example
that we studied, it was found that the rate asymmetries for these
decay modes are particularly sensitive to the phase of $A_t$, although
other phases contribute to the asymmetries as well.  For the
parameters considered, asymmetries of order $10\%$ were found for
${\tilde\chi}^\pm_2 \to {\tilde\chi}^\pm_1 H_1H_1$; somewhat smaller
asymmetries were found for the other decay modes.  Should SUSY be
discovered, chargino decays could provide a useful avenue for
investigating the CP structure of the theory.


\begin{acknowledgments}
We would like to thank J.S.~Lee for helpful correspondence.  This work
was financially supported by NSERC of Canada. The work of J.H.,
K.K. and K.L. was supported in part by the U.S.\ National Science
Foundation under Grant PHY--0601103; K.K. was also supported in part
by the U.S.\ National Science Foundation under Grant PHY--0900914.
\end{acknowledgments}


\appendix*

\section{Expressions for the Decay Amplitudes of 
${\widetilde \chi}_2^{\pm} \to {\widetilde \chi}_1^{\pm} XY$}

This Appendix contains analytical expressions for the amplitudes
associated with the processes considered in this paper.

Let us first clarify our notation for the various propagators.  We
define Breit-Wigner-type propagators for chargino and neutralino
internal lines as follows,
\bea 
   i (\slash\!\!\!p + m) \tilde{D}(p^2,m^2,\Gamma)
   \equiv \frac{i (\slash\!\!\!p + m)}{p^2 - m^2 + i\Gamma m}
\eea 
where the tilde is used to distinguish the Breit-Wigner propagators
from the $3\times 3$ Higgs propagator $D_{ij}$ to be described below.
We also employ the Breit-Wigner form of the propagator for the graphs
mediated by a $Z$ boson. The $Z$ propagator in the unitary gauge is $i
\tilde{D}(p^2_Z,m_Z^2,\Gamma_Z) (-g^{\alpha \beta} + p_Z^\alpha
p_Z^\beta/m_Z^2)$.\footnote{There has been some discussion in the
  literature regarding the correct form to use for the propagator of a
  spin-1 particle when the particle's width contributes to a rate
  asymmetry.  See, for example, Ref.~\cite{atwood1994}.  While this
  discussion is important in some contexts, we nevertheless use the
  ``naive'' form of the $Z$ propagator in our numerical work, since
  the intermediate $Z$ boson is far off-shell in the examples we
  consider.}

As noted in the text, our calculation employs the $3\times 3$ physical
Higgs boson block of the full $4\times4$ neutral Higgs propagator
computed by {\tt CPsuperH2.0}.  We also differ notationally from {\tt
  CPsuperH2.0} in terms of the over-all normalization of the
propagator.  The specific relationship between the two sets of
notation (within the physical $3\times3$ block) is the following,
\bea
   D_{ij}(M^2)=D_{ij}^{\mbox{\scriptsize CPsuperH}}(M^2)/M^2\; .
\eea

Unless noted otherwise, our notation for coupling constants and
diagonalization matrices follows the notation used in
Ref.~\cite{CPsuperH}.  One exception is the definition of
$g_{H_iH_jZ}$ (which occurs in the ``$ZH_1$'' decay), for which we use
the notation defined in Ref.~\cite{previous2}.

In the following, we include explicit expressions for the ${\widetilde
  \chi}_2^-$ decays.  The corresponding expressions for the
${\widetilde \chi}_2^+$ decays can be obtained from the given
expressions by taking the complex conjugates of the Lorentz-invariant
pieces ($B$, $C$, $D$...), with the exception of keeping the
propagator functions $D_{ij}$ and $\tilde{D}$ unchanged. As an
example, this procedure is demonstrated explicitly for the case of
${\widetilde \chi}_2 \to {\widetilde \chi}_1 H_1 H_1$.

\subsection{\boldmath $X,Y=H_1H_1$}
The amplitude for ${\widetilde \chi}_2^- \to {\widetilde \chi}_1^- H_1
H_1$ is given by
\begin{eqnarray} 
{\cal M}^{H_1 H_1}&=&
\overline{u}_{\widetilde\chi_1}(s_2,p_2)\left[\left(B+C\gamma^5\right)+
\slash\!\!\!\xi\left(D+F\gamma^5\right)+\slash\!\!\!\rho\left(G+H\gamma^5
\right)\right]u_{\widetilde\chi_2}(s_1,p_1)  
\,,\nonumber \\
\label{eq:a1}
\end{eqnarray}
where $\rho^\mu = (p_2+p_3)^\mu$, $\xi^\mu = (p_2+p_4)^\mu$ and 
\begin{eqnarray} 
B&=&\frac{gv}{\sqrt{2}}\!\sum_{i,j=1}^{3}\!g^{S}_{H_i\widetilde\chi_1^+
\widetilde\chi_2^-}
D_{ij}(M^2)g_{H_jH_1H_1}\eta_j\\
&&-\frac{g^2}{2}\!\sum_{k=1}^{2}\eta_{SP,k}^{(-)}~m_{\widetilde\chi_k}\!
\left[\tilde{D}(\xi^2,m_{\widetilde\chi_k}^2,\Gamma_{\widetilde\chi_k})+\tilde{D}(\rho^2,m_{\widetilde\chi_k}^2,\Gamma_{\widetilde\chi_k})\right]\!,\nonumber
\\
C&=&\frac{igv}{\sqrt{2}}\!\sum_{i,j=1}^{3}\!g^{P}_{H_i\widetilde\chi_1^+
\widetilde\chi_2^-}
D_{ij}(M^2)g_{H_jH_1H_1}\eta_j\\
&&-\frac{ig^2}{2}\!\sum_{k=1}^{2}\eta_{PS,k}^{(+)}m_{\widetilde\chi_k}\!\!
\left[\tilde{D}(\xi^2,m_{\widetilde\chi_k}^2,\Gamma_{\widetilde\chi_k})+\tilde{D}(\rho^2,m_{\widetilde\chi_k}^2,\Gamma_{\widetilde\chi_k})\right]\!,\nonumber
\\
D&=&-\frac{g^2}{2}\sum_{k=1}^{2}\eta_{SP,k}^{(+)}\,\tilde{D}(\xi^2,m_{\widetilde\chi_k}^2,\Gamma_{\widetilde\chi_k})\,,\nonumber
\\
F&=&-\frac{ig^2}{2}\sum_{k=1}^{2}\eta_{PS,k}^{(-)}\,\tilde{D}(\xi^2,m_{\widetilde\chi_k}^2,\Gamma_{\widetilde\chi_k})\,,\nonumber
\\
G&=&-\frac{g^2}{2}\sum_{k=1}^{2}\eta_{SP,k}^{(+)}\,\tilde{D}(\rho^2,m_{\widetilde\chi_k}^2,\Gamma_{\widetilde\chi_k})\,,\nonumber
\\
H&=&-\frac{ig^2}{2}\sum_{k=1}^{2}\eta_{PS,k}^{(-)}\,\tilde{D}(\rho^2,m_{\widetilde\chi_k}^2,\Gamma_{\widetilde\chi_k})\,,
\end{eqnarray}
where we have defined $M^2=(p_3+p_4)^2$, $\eta^{(\pm)}_{\alpha\beta,k}
=g^S_{H_1\widetilde\chi_1^+\widetilde\chi_k^-}
g^\alpha_{H_1\widetilde\chi_k^+\widetilde\chi_2^-} \pm
g^P_{H_1\widetilde\chi_1^+\widetilde\chi_k^-}
g^\beta_{H_1\widetilde\chi_k^+\widetilde\chi_2^-}$, and the factors
$\eta_j$ are $\eta_1=6$ and $\eta_{2,3}=2$.
Also, the $m_{\widetilde\chi_k}$ denote the chargino masses.

The amplitude for ${\widetilde \chi}_2^+ \to {\widetilde \chi}_1^+ H_1
H_1$ is given by
\begin{eqnarray} 
{\bar{\cal M}}^{H_1 H_1}&=&
\overline{u}_{\widetilde\chi_1}(s_2,p_2)\left[\left(\bar B+\bar
C\gamma^5\right)+ \slash\!\!\!\xi\left(\bar D+\bar
F\gamma^5\right)+\slash\!\!\!\rho\left(\bar G+\bar H\gamma^5
\right)\right]u_{\widetilde\chi_2}(s_1,p_1)   \,,\nonumber \\
\label{eq:a3}
\end{eqnarray}
where
\begin{eqnarray} 
\bar B&=&\frac{gv}{\sqrt{2}}\sum_{i,j=1}^{3}g^{S*}_{H_i\widetilde\chi_1^+\widetilde\chi_2^-}
D_{ij}(M^2)g_{H_jH_1H_1}\eta_j\nonumber\\
& & -\frac{g^2}{2}\sum_{k=1}^{2}\eta_{SP,k}^{(-)*}~m_{\widetilde\chi_k}
\left[\tilde{D}(\xi^2,m_{\widetilde\chi_k}^2,\Gamma_{\widetilde\chi_k})+\tilde{D}(\rho^2,m_{\widetilde\chi_k}^2,\Gamma_{\widetilde\chi_k})\right]\,,\nonumber \\
\bar C&=&-\frac{igv}{\sqrt{2}}\sum_{i,j=1}^{3}g^{P*}_{H_i\widetilde\chi_1^+\widetilde\chi_2^-}
D_{ij}(M^2)g_{H_jH_1H_1}\eta_j\nonumber \\
& & +\frac{ig^2}{2}\sum_{k=1}^{2}\eta_{PS,k}^{(+)*}~m_{\widetilde{\chi_k}}
\left[\tilde{D}(\xi^2,m_{\widetilde\chi_k}^2,\Gamma_{\widetilde\chi_k})+\tilde{D}(\rho^2,m_{\widetilde\chi_k}^2,\Gamma_{\widetilde\chi_k})\right]\,,\nonumber \\
\bar D&=&-\frac{g^2}{2}\sum_{k=1}^{2}\eta_{SP,k}^{(+)*}\,\tilde{D}(\xi^2,m_{\widetilde\chi_k}^2,\Gamma_{\widetilde\chi_k})\,,\nonumber \\
\bar F&=&\frac{ig^2}{2}\sum_{k=1}^{2}\eta_{PS,k}^{(-)*}\,\tilde{D}(\xi^2,m_{\widetilde\chi_k}^2,\Gamma_{\widetilde\chi_k})\,,\nonumber \\
\bar G&=&-\frac{g^2}{2}\sum_{k=1}^{2}\eta_{SP,k}^{(+)*}\,\tilde{D}(\rho^2,m_{\widetilde\chi_k}^2,\Gamma_{\widetilde\chi_k})\,,\nonumber \\
\bar H&=&\frac{ig^2}{2}\sum_{k=1}^{2}\eta_{PS,k}^{(-)*}\,\tilde{D}(\rho^2,m_{\widetilde\chi_k}^2,\Gamma_{\widetilde\chi_k})\,.
\end{eqnarray}
Note that the spinors $u_{\widetilde\chi_{1,2}}$ in Eqs.~(\ref{eq:a1})
and (\ref{eq:a3}) actually refer to $u_{\widetilde\chi^-_{1,2}}$ in
both cases.  Here, and throughout this work, we have manipulated
expressions in such a way that the ${\widetilde\chi}_2^+$ decay
amplitudes are written in terms of the spinors
$u_{\widetilde\chi^-_{1,2}}$.
\subsection{\boldmath $X,Y=ZZ$}

The amplitude for ${\widetilde \chi}_2^- \to {\widetilde \chi}_1^- ZZ$
is given by
\begin{eqnarray} 
{\cal M}^{ZZ}&=&\overline{u}_{\widetilde\chi_1}(s_2,p_2)\left[\left(B+C\gamma^5\right)g^{\mu\nu}+
\gamma^{\nu}\slash\!\!\!\xi\gamma^{\mu}\left(D+F\gamma^5\right)\right. \nonumber \\
& &\left. +\gamma^{\mu}\slash\!\!\!\rho\gamma^{\nu}\left(G+H\gamma^5\right)+
\gamma^{\mu}\gamma^{\nu}\left(J+K\gamma^5\right) \right]u_{\widetilde\chi_2}(s_1,p_1)
\epsilon^{\lambda_1*}_{\mu}\epsilon^{\lambda_2*}_{\nu}\,, 
\end{eqnarray}
where 
\begin{eqnarray} 
  B&=&\frac{g^2m_W}{\sqrt{2}\cos^2\theta_W}\sum_{i,j=1}^{3}g^{S}_{H_i\widetilde\chi_1^+\widetilde\chi_2^-}
  D_{ij}(M^2)g_{H_jVV}\\
&&-\frac{g^2}{4\cos^2\theta_W}\sum_{k=1}^{2}
  m_{\widetilde\chi_k}\,\omega^+_{RL,k}\,\tilde{D}(\xi^2,m_{\widetilde\chi_k}^2,\Gamma_{\widetilde\chi_k})\,,\nonumber \\
  C&=&\frac{ig^2m_W}{\sqrt{2}\cos^2\theta_W}\sum_{i,j=1}^{3}g^{P}_{H_i\widetilde\chi_1^+\widetilde\chi_2^-}
  D_{ij}(M^2)g_{H_jVV}\\
&&-\frac{g^2}{4\cos^2\theta_W}\sum_{k=1}^{2}
  m_{\widetilde\chi_k}\,\omega^-_{RL,k}\,\tilde{D}(\xi^2,m_{\widetilde\chi_k}^2,\Gamma_{\widetilde\chi_k})\,,\nonumber \\
  D&=&-\frac{g^2}{8\cos^2\theta_W}\sum_{k=1}^{2}\omega^+_{LR,k}\,\tilde{D}(\xi^2,m_{\widetilde\chi_k}^2,\Gamma_{\widetilde\chi_k})\,, \nonumber \\
  F&=&\frac{g^2}{8\cos^2\theta_W}\sum_{k=1}^{2}\omega^-_{LR,k}\,\tilde{D}(\xi^2,m_{\widetilde\chi_k}^2,\Gamma_{\widetilde\chi_k})\,, \nonumber \\
  G&=&-\frac{g^2}{8\cos^2\theta_W}\sum_{k=1}^{2}\omega^+_{LR,k}\,\tilde{D}(\rho^2,m_{\widetilde\chi_k}^2,\Gamma_{\widetilde\chi_k})\,, \nonumber \\
  H&=&\frac{g^2}{8\cos^2\theta_W}\sum_{k=1}^{2}\omega^-_{LR,k}\,\tilde{D}(\rho^2,m_{\widetilde\chi_k}^2,\Gamma_{\widetilde\chi_k})\,, \nonumber \\
  J&=&\sum_{k=1}^{2}\frac{g^2\,m_{\widetilde\chi_k}\,\omega^+_{RL,k}}{8\cos^2\theta_W} 
  \left[\tilde{D}(\xi^2,m_{\widetilde\chi_k}^2,\Gamma_{\widetilde\chi_k})-\tilde{D}(\rho^2,m_{\widetilde\chi_k}^2,\Gamma_{\widetilde\chi_k})\right]\,,\nonumber \\
  K&=&\sum_{k=1}^{2}\frac{g^2\,m_{\widetilde\chi_k}\,\omega^-_{RL,k}}{8\cos^2\theta_W}
  \left[\tilde{D}(\xi^2,m_{\widetilde\chi_k}^2,\Gamma_{\widetilde\chi_k})-\tilde{D}(\rho^2,m_{\widetilde\chi_k}^2,\Gamma_{\widetilde\chi_k})\right]\,,
\end{eqnarray}
and where we have defined
$\omega^{(\pm)}_{\alpha\beta,k}=
U^{1k}_{(L)}U^{k2}_{(\alpha)} \pm U^{1k}_{(R)}U^{k2}_{(\beta)}$, with
\begin{eqnarray}
U_{(L)}^{ij}&=&
(C_L)^{i1}(C_L)^{\ast j1}+(\cos^2\theta_W-\sin^2\theta_W)\delta^{ij}\,,
\nonumber \\
U_{(R)}^{ij}&=&
(C_R)^{i1}(C_R)^{\ast j1}+(\cos^2\theta_W-\sin^2\theta_W)\delta^{ij}\,.
\end{eqnarray} 
The unitary matrices $C_L$ and $C_R$ are used to diagonalize the
chargino mass matrix.  The matrices $U_{(L,R)}$ appear in the
chargino-chargino-$Z$ Lagrangian as follows,
\begin{eqnarray}
{\cal{L}}_{Z\chi_i\chi_j}=\frac{g}{2\cos\theta_W}
\overline{\chi^-_i} \gamma^\mu \bigl[U_{(L)}^{ij}P_L+U_{(R)}^{ij}P_R\bigr]\chi^-_j Z_\mu,
\end{eqnarray} 
where $P_{L(R)}=(1-(+)\gamma^5)/2$.
%


\subsection{\boldmath $X,Y=W^+W^-$}

The amplitude for ${\widetilde \chi}_2^- \to {\widetilde \chi}_1^-
W^+W^-$ is given by
\begin{eqnarray} 
{\cal M}^{WW}&=& \overline{u}_{\widetilde\chi_1}(s_2,p_2)\left[\left(B+C\gamma^5\right)g^{\mu\nu}+
\gamma^{\mu}\slash\!\!\!\rho\gamma^{\nu}\left(D+F\gamma^5\right)+\gamma^{\mu}\gamma^{\nu}
\left(G+H\gamma^5\right) \right.  \nonumber \\
& &\!\!\!\!\!\!\left. +\left(\gamma^{\nu}p_4^{\mu}-\gamma^{\mu}p_3^{\nu}+\frac{1}{2}(\slash\!\!\!p_{3}-\slash\!\!\!p_{4})\,g^{\mu\nu}
\right)\left(J+K\gamma^5\right)
\right]u_{\widetilde\chi_2}(s_1,p_1) \epsilon^{\lambda_1*}_{\mu}\epsilon^{\lambda_2*}_{\nu}\!, 
\end{eqnarray}
where $p_4$ denotes the momentum of the $W^-$ and where
\begin{eqnarray} 
  B&=&\frac{g^2m_W}{\sqrt{2}}\sum_{i,j=1}^{3}g^{S}_{H_i\widetilde\chi_1^+\widetilde\chi_2^-}~g_{H_jVV}
  D_{ij}(M^2) \,,\nonumber \\
  C&=&\frac{ig^2m_W}{\sqrt{2}}\sum_{i,j=1}^{3}g^{P}_{H_i\widetilde\chi_1^+\widetilde\chi_2^-}~g_{H_jVV}
  D_{ij}(M^2)\,, \nonumber \\
  D&=&-\frac{g^2}{2}\,\sum_{k=1}^{4}\bar{V}^+_{LR,k}\,\tilde{D}(\rho^2,m_{\widetilde\chi_k^0}^2,\Gamma_{\widetilde\chi_k^0})\,, \nonumber \\
  F&=&\frac{g^2}{2}\,\sum_{k=1}^{4}\bar{V}^-_{LR,k}\,\tilde{D}(\rho^2,m_{\widetilde\chi_k^0}^2,\Gamma_{\widetilde\chi_k^0})\,, \nonumber \\
  G&=&-\frac{g^2}{2}\,\sum_{k=1}^{4}m_{\widetilde\chi_k^0}\bar{V}^+_{RL,k}\,\tilde{D}(\rho^2,m_{\widetilde\chi_k^0}^2,\Gamma_{\widetilde\chi_k^0})\,, \nonumber \\
  H&=&-\frac{g^2}{2}\,\sum_{k=1}^{4}m_{\widetilde\chi_k^0}\bar{V}^-_{RL,k}\,\tilde{D}(\rho^2,m_{\widetilde\chi_k^0}^2,\Gamma_{\widetilde\chi_k^0})\,,\nonumber \\
  J&=&\frac{g^2}{2}\left[U^{12}_{R}+U^{12}_{L}\right]\tilde{D}(M^2,m_Z^2,\Gamma_Z)\,, \nonumber \\
  K&=&\frac{g^2}{2}\left[U^{12}_{R}-U^{12}_{L}\right]\tilde{D}(M^2,m_Z^2,\Gamma_Z)\,.
\end{eqnarray}
In the above expressions, $m_{\widetilde\chi_k^0}$ and
$\Gamma_{\widetilde\chi_k^0}$ denote the neutralino masses and widths,
respectively.  Also, we define $\bar{V}_{\alpha\beta,k}^{\pm}$ as
$\bar{V}_{\alpha\beta,k}^{\pm}=V_{(L)}^{k1}V_{(\alpha)}^{k2*}\pm
V_{(R)}^{k1}V_{(\beta)}^{k2*}$, where
\begin{eqnarray} 
V_{(L)}^{ij}&=&
(N)^{\ast i2}(C_L)^{j1}+
\frac{1}{\sqrt{2}}(N)^{\ast i3}(C_L)^{j2}\,,
\nonumber \\
V_{(R)}^{ij}&=&
(N)^{i2}(C_R)^{j1}-
\frac{1}{\sqrt{2}}(N)^{i4}(C_R)^{j2}\,.
\nonumber
\end{eqnarray}
The $4\times 4$ unitary matrix $N$ is used to diagonalize the
neutralino mass matrix.


\subsection{\boldmath $X,Y=Z H_1$}

The amplitude for ${\tilde \chi}_2^- \to {\tilde \chi}_1^- ZH_1$ is given by
\begin{eqnarray} 
{\cal M}^{ZH_1}&=&\overline{u}_{\widetilde\chi_1}(s_2,p_2)\left[\left(B+C\gamma^5\right)p_4^{\mu}+
\left(D+F\gamma^5\right)\gamma^{\mu}+\left(G+H\gamma^5\right)\slash\!\!\!\xi\gamma^{\mu}\right. \nonumber \\
& &\left. +\left(J+K\gamma^5\right)\gamma^{\mu}\slash\!\!\!\rho\right]u_{\widetilde\chi_2}(s_1,p_1)
\epsilon_{\mu}^{\lambda*}\,,
\end{eqnarray}
where $p_4$ denotes the momentum of the $H_1$ in the final state and
where
\begin{eqnarray} 
 B&=&-\frac{ig^2}{\sqrt{2}\cos\theta_W}\sum_{i=1,~j=2}^{3}
  g^{S}_{H_i\widetilde\chi^+_1\widetilde\chi^-_2}D_{ij}(M^2)~g_{H_1H_jZ}\nonumber\\
  & & -\frac{g^2m_W}{8\cos^{3}\theta_Wm^2_Z}
  g_{H_1VV}\left(m_{\widetilde{\chi_2}}-m_{\widetilde{\chi}_1}\right)
        \left[U_{L}^{12}+U_{R}^{12}\right]\tilde{D}(M^2,m_Z^2,\Gamma_Z)
  \,, \nonumber \\
 C&=&\frac{g^2}{\sqrt{2}\cos\theta_W}\sum_{i=1,~j=2}^{3}
  g^{P}_{H_i\widetilde\chi^+_1\widetilde\chi^-_2}D_{ij}(M^2)~g_{H_1H_jZ}\nonumber \\
  & & -\frac{g^2m_W}{8\cos^{3}\theta_W m^2_Z}
  g_{H_1VV}\left(m_{\widetilde{\chi}_1}+m_{\widetilde{\chi}_2}\right)
       \left[U_{L}^{12}-U_{R}^{12}\right]\tilde{D}(M^2,m_Z^2,\Gamma_Z)
  \,, \nonumber \\
 D&=&\!\sum_{k=1}^{2}\!\frac{g^2 m_{\widetilde\chi_k}\tilde{D}(\rho^2,m_{\widetilde\chi_k}^2,\Gamma_{\widetilde\chi_k})}
  {4\sqrt{2}\cos\theta_W}\!\left[
  U^{1k}_L(g_{H_1\widetilde\chi_k^+\widetilde\chi_2^-}^{S}\!-\!ig_{H_1\widetilde\chi_k^+\widetilde\chi_2^-}^{P})
 \!+\!U^{1k}_R(g_{H_1\widetilde\chi_k^+\widetilde\chi_2^-}^{S}\!+\!ig_{H_1\widetilde\chi_k^+\widetilde\chi_2^-}^{P}
  )\right]\nonumber \\
  & + &\!\sum_{k=1}^{2}\!\frac{g^2 m_{\widetilde\chi_k}\tilde{D}(\xi^2,m_{\widetilde\chi_k}^2,\Gamma_{\widetilde\chi_k})}
  {4\sqrt{2}\cos\theta_W}\!\left[
  U^{k2}_L(g_{H_1\widetilde\chi_1^+\widetilde\chi_k^-}^{S}\!+\!ig_{H_1\widetilde\chi_1^+\widetilde\chi_k^-}^{P})
 \!+\!U^{k2}_R(g_{H_1\widetilde\chi_1^+\widetilde\chi_k^-}^{S}\!-\!ig_{H_1\widetilde\chi_1^+\widetilde\chi_k^-}^{P})
  \right] \nonumber \\
  &+ &\!\!\frac{g^2m_W}{8\cos^{3}\theta_W}g_{H_1VV}
  \left[U_L^{12}+U_R^{12}\right]\tilde{D}(M^2,m_Z^2,\Gamma_Z) ,\nonumber \\
 F&=&\!\sum_{k=1}^{2}\!\frac{g^2 m_{\widetilde\chi_k} \tilde{D}(\rho^2,m_{\widetilde\chi_k}^2,\Gamma_{\widetilde\chi_k})}{4\sqrt{2}\cos\theta_W}
  \!\left[
  U^{1k}_L(g_{H_1\widetilde\chi_k^+\widetilde\chi_2^-}^{S}\!-\!ig_{H_1\widetilde\chi_k^+\widetilde\chi_2^-}^{P})
 \!-\!U^{1k}_R(g_{H_1\widetilde\chi_k^+\widetilde\chi_2^-}^{S}\!+\!ig_{H_1\widetilde\chi_k^+\widetilde\chi_2^-}^{P})
  \right] \nonumber \\
  &+ &\!\sum_{k=1}^{2}\!\frac{g^2 m_{\widetilde\chi_k} \tilde{D}(\xi^2,m_{\widetilde\chi_k}^2,\Gamma_{\widetilde\chi_k})}
  {4\sqrt{2}\cos\theta_W}\!\left[
  U^{k2}_L(g_{H_1\widetilde\chi_1^+\widetilde\chi_k^-}^{S}\!+\!ig_{H_1\widetilde\chi_1^+\widetilde\chi_k^-}^{P})
 \!-\!U^{k2}_R(g_{H_1\widetilde\chi_1^+\widetilde\chi_k^-}^{S}\!-\!ig_{H_1\widetilde\chi_1^+\widetilde\chi_k^-}^{P}) 
  \right]\nonumber \\
  &+ &\!\!\frac{g^2m_W}{8\cos^{3}\theta_W}g_{H_1VV}
  \left[U_L^{12}-U_R^{12}\right] \tilde{D}(M^2,m_Z^2,\Gamma_Z)\,,\nonumber \\
 G&=&\!\sum_{k=1}^{2}\!\frac{g^2 \tilde{D}(\xi^2,m_{\widetilde\chi_k}^2,\Gamma_{\widetilde\chi_k})}{4\sqrt{2}\cos\theta_W}\!
  \left[
  U^{k2}_L(g_{H_1\widetilde\chi_1^+\widetilde\chi_k^-}^{S}\!-\!ig_{H_1\widetilde\chi_1^+\widetilde\chi_k^-}^{P})
 \!+\!U^{k2}_R(g_{H_1\widetilde\chi_1^+\widetilde\chi_k^-}^{S}\!+\!ig_{H_1\widetilde\chi_1^+\widetilde\chi_k^-}^{P})  
  \right] \!, \nonumber \\
 H&=&\!\sum_{k=1}^{2}\!\frac{g^2 \tilde{D}(\xi^2,m_{\widetilde\chi_k}^2,\Gamma_{\widetilde\chi_k})}{4\sqrt{2}\cos\theta_W}\!
  \left[
 \!-U^{k2}_L(g_{H_1\widetilde\chi_1^+\widetilde\chi_k^-}^{S}\!-\!ig_{H_1\widetilde\chi_1^+\widetilde\chi_k^-}^{P})
 \!+\!U^{k2}_R(g_{H_1\widetilde\chi_1^+\widetilde\chi_k^-}^{S}\!+\!ig_{H_1\widetilde\chi_1^+\widetilde\chi_k^-}^{P})\!  
  \right] \!, \nonumber \\
 J&=&\!\sum_{k=1}^{2}\!\frac{g^2 \tilde{D}(\rho^2,m_{\widetilde\chi_k}^2,\Gamma_{\widetilde\chi_k})}{4\sqrt{2}\cos\theta_W}\!
  \left[
  U^{1k}_L(g_{H_1\widetilde\chi_k^+\widetilde\chi_2^-}^{S}\!+\!ig_{H_1\widetilde\chi_k^+\widetilde\chi_2^-}^{P})
 +U^{1k}_R(g_{H_1\widetilde\chi_k^+\widetilde\chi_2^-}^{S}\!-\!ig_{H_1\widetilde\chi_k^+\widetilde\chi_2^-}^{P})\!
  \right]\!, \nonumber \\
 K&=&\!\sum_{k=1}^{2}\!\frac{g^2 \tilde{D}(\rho^2,m_{\widetilde\chi_k}^2,\Gamma_{\widetilde\chi_k})}{4\sqrt{2}\cos\theta_W}\!
  \left[
  U^{1k}_L(g_{H_1\widetilde\chi_k^+\widetilde\chi_2^-}^{S}\!+ig_{H_1\widetilde\chi_k^+\widetilde\chi_2^-}^{P})
 \!-\!U^{1k}_R(g_{H_1\widetilde\chi_k^+\widetilde\chi_2^-}^{S}\!-ig_{H_1\widetilde\chi_k^+\widetilde\chi_2^-}^{P})\!
  \right]\!.\nonumber\\
\end{eqnarray}
As noted above, our notation for $g_{H_iH_jZ}$ is the same as that adopted
in Ref.~\cite{previous2}.



\begin{thebibliography}{99}

\bibitem{previous1} 
  K.~Kiers, A.~Szynkman and D.~London,
  Phys.\ Rev.\  D {\bf 74}, 035004 (2006)
  [arXiv:hep-ph/0605123].

\bibitem{previous2} 
  A.~Szynkman, K.~Kiers and D.~London,
  Phys.\ Rev.\  D {\bf 75}, 075009 (2007)
  [arXiv:hep-ph/0701165].

\bibitem{yang}
  W.~M.~Yang and D.~S.~Du,
  Phys.\ Rev.\  D {\bf 67}, 055004 (2003)
  [arXiv:hep-ph/0211453].

\bibitem{bartl2004}
  A.~Bartl, H.~Fraas, O.~Kittel and W.~Majerotto,
  Phys.\ Lett.\  B {\bf 598}, 76 (2004)
  [arXiv:hep-ph/0406309].

\bibitem{kittel}
  O.~Kittel, A.~Bartl, H.~Fraas and W.~Majerotto,
  Phys.\ Rev.\  D {\bf 70}, 115005 (2004)
  [arXiv:hep-ph/0410054].

\bibitem{eberl}
  H.~Eberl, T.~Gajdosik, W.~Majerotto and B.~Schrausser,
  Phys.\ Lett.\  B {\bf 618}, 171 (2005)
  [arXiv:hep-ph/0502112].

\bibitem{bartl2006}
  A.~Bartl, H.~Fraas, S.~Hesselbach, K.~Hohenwarter-Sodek, T.~Kernreiter and G.~Moortgat-Pick,
  Eur.\ Phys.\ J.\  C {\bf 51}, 149 (2007)
  [arXiv:hep-ph/0608065].

\bibitem{osland}
  P.~Osland and A.~Vereshagin,
  Phys.\ Rev.\  D {\bf 76}, 036001 (2007)
  [arXiv:0704.2165 [hep-ph]].

\bibitem{rolbiecki}
  K.~Rolbiecki and J.~Kalinowski,
  Phys.\ Rev.\  D {\bf 76}, 115006 (2007)
  [arXiv:0709.2994 [hep-ph]].

\bibitem{bartl}
  A.~Bartl, K.~Hohenwarter-Sodek, T.~Kernreiter, O.~Kittel and M.~Terwort,
  Nucl.\ Phys.\  B {\bf 802}, 77 (2008)
  [arXiv:0802.3592 [hep-ph]].

\bibitem{Kittel:2008be}
  O.~Kittel and F.~von der Pahlen,
  JHEP {\bf 0808}, 030 (2008)
  [arXiv:0806.4534 [hep-ph]].


\bibitem{mixing} F.~Gabbiani and A.~Masiero, Nucl.\ Phys.\ B
{\bf 322}, 235 (1989);
S.~Bertolini, F.~Borzumati, A.~Masiero and G.~Ridolfi, Nucl.\
Phys.\ B {\bf 353}, 591 (1991);
J.~S.~Hagelin, S.~Kelley and T.~Tanaka, Nucl.\ Phys.\ B {\bf
  415}, 293 (1994).

\bibitem{B-meson} A.~Ali and D.~London, Eur.\ Phys.\ J.\ C
{\bf 9}, 687 (1999).

\bibitem{edms} F.~Gabbiani, E.~Gabrielli, A.~Masiero and
L.~Silvestrini, Nucl.\ Phys.\ B {\bf 477}, 321 (1996);
Y.~Grossman, Y.~Nir and R.~Rattazzi, Adv.\ Ser.\ Direct.\
High Energy Phys.\ {\bf 15}, 755 (1998);
D.~Chang, W-Y.~Keung and A.~Pilaftsis, Phys.\ Rev.\ Lett.\
{\bf 82} 900 (1999);
A.~Pilaftsis and C.~E.~M.~Wagner, Nucl.\ Phys.\ B {\bf 553},
3 (1999);
S.~Abel, S.~Khalil and O.~Lebedev, Nucl.\ Phys.\ B {\bf 606},
151 (2001);
D.~A.~Demir, O.~Lebedev, K.~A.~Olive, M.~Pospelov and
A.~Ritz, Nucl.\ Phys.\ B {\bf 680}, 339 (2004);
K.~A.~Olive, M.~Pospelov, A.~Ritz and Y.~Santoso, Phys.\
Rev.\ D {\bf 72}, 075001 (2005).

\bibitem{CPsuperH} J.~S.~Lee, A.~Pilaftsis, M.~Carena,
S.~Y.~Choi, M.~Drees, J.~R.~Ellis and C.~E.~M.~Wagner,
Comput.\ Phys.\ Commun. {\bf 156}, 283 (2004).

\bibitem{CPsuperH2a}
  J.~S.~Lee, M.~Carena, J.~Ellis, A.~Pilaftsis and C.~E.~M.~Wagner,
  Comput.\ Phys.\ Commun.\  {\bf 180}, 312 (2009)
  [arXiv:0712.2360 [hep-ph]].

\bibitem{CPsuperH2b}
  J.~R.~Ellis, J.~S.~Lee and A.~Pilaftsis,
  JHEP {\bf 0810}, 049 (2008)
  [arXiv:0808.1819 [hep-ph]].

\bibitem{widths} J.~R.~Ellis, J.~S.~Lee and A.~Pilaftsis,
Phys.\ Rev.\ D {\bf 70}, 075010 (2004).

\bibitem{higgs_tree} See, for example, J.F. Gunion,
H.E. Haber, G.L. Kane and S. Dawson, ``The Higgs Hunter's
Guide,'' (Addison-Wesley, Reading, MA, 1990) and H.E. Haber,
arXiv:hep-ph/9707213.

\bibitem{rosiek1}
  J.~Rosiek,
  Phys.\ Rev.\  D {\bf 41}, 3464 (1990).

\bibitem{rosiek2}
  J.~Rosiek,
  arXiv:hep-ph/9511250.

\bibitem{feyncalc}
  R.~Mertig, M.~Bohm and A.~Denner,
  Comput.\ Phys.\ Commun.\  {\bf 64}, 345 (1991).

\bibitem{gerardhou}
  J.~M.~Gerard and W.~S.~Hou,
  Phys.\ Rev.\ Lett.\  {\bf 62}, 855 (1989).

\bibitem{wolfenstein}
  L.~Wolfenstein,
  Phys.\ Rev.\  D {\bf 43}, 151 (1991).

\bibitem{pinch1}
  K.~Philippides and A.~Sirlin,
  Phys.\ Lett.\  B {\bf 367}, 377 (1996)
  [arXiv:hep-ph/9510393].

\bibitem{pinch2}
  J.~Papavassiliou and A.~Pilaftsis,
  Phys.\ Rev.\  D {\bf 53}, 2128 (1996)
  [arXiv:hep-ph/9507246].

\bibitem{CPXoriginal}
  M.~S.~Carena, J.~R.~Ellis, A.~Pilaftsis and C.~E.~M.~Wagner,
  Phys.\ Lett.\  B {\bf 495}, 155 (2000)
  [arXiv:hep-ph/0009212].

\bibitem{ginges2003}
  J.~S.~M.~Ginges and V.~V.~Flambaum,
  Phys.\ Rept.\  {\bf 397}, 63 (2004)
  [arXiv:physics/0309054].

\bibitem{regan2002}
  B.~C.~Regan, E.~D.~Commins, C.~J.~Schmidt and D.~DeMille,
  Phys.\ Rev.\ Lett.\  {\bf 88}, 071805 (2002).

\bibitem{griffith2009}
  W.~C.~Griffith, M.~D.~Swallows, T.~H.~Loftus, M.~V.~Romalis, 
  B.~R.~Heckel and E.~N.~Fortson,
  Phys.\ Rev.\ Lett.\ {\bf 102}, 101601 (2009)
  [arXiv:0901.2328 [physics.atom-ph]].

\bibitem{baker2006}
  C.~A.~Baker {\it et al.},
  Phys.\ Rev.\ Lett.\  {\bf 97}, 131801 (2006)
  [arXiv:hep-ex/0602020].

\bibitem{bennett}
  G.~W.~Bennett {\it et al.}  [Muon (g-2) Collaboration],
  Phys.\ Rev.\  D {\bf 80}, 052008 (2009)
  [arXiv:0811.1207 [hep-ex]].

\bibitem{cheung2009}
  K.~Cheung, O.~C.~W.~Kong and J.~S.~Lee,
  JHEP {\bf 0906}, 020 (2009)
  [arXiv:0904.4352 [hep-ph]].

\bibitem{hfag}
 Heavy Flavor Averaging Group, {\tt http://www.slac.stanford.edu/xorg/hfag}.

\bibitem{pdg}
  C.~Amsler {\it et al.}  [Particle Data Group],
  Phys.\ Lett.\ B {\bf 667}, 1 (2008)
  and 2009 partial update for the 2010 edition (URL: {\tt
    http://pdg.lbl.gov}).

\bibitem{ellis2007}
  J.~R.~Ellis, J.~S.~Lee and A.~Pilaftsis,
  Phys.\ Rev.\  D {\bf 76}, 115011 (2007)
  [arXiv:0708.2079 [hep-ph]].

\bibitem{atwood1994}
  D.~Atwood, G.~Eilam, A.~Soni, R.~R.~Mendel and R.~Migneron,
  Phys.\ Rev.\  D {\bf 49}, 289 (1994).


\end{thebibliography}
\end{document}